\documentclass{jpp}
\pdfoutput=1

\usepackage[T1]{fontenc}
\usepackage[utf8]{inputenc}
\usepackage[english]{babel} 
\usepackage{amsmath}
\usepackage{amssymb}
\usepackage{textcomp}
\usepackage{lmodern}

\usepackage{color}
\usepackage{graphicx}
\usepackage[hidelinks=true,
colorlinks=true,
linkcolor=blue,
citecolor=blue,
urlcolor=blue
]{hyperref}
\usepackage{natbib}

\graphicspath{ {figures/} }

\usepackage{physics}
\usepackage[nosfdefault]{comicneue}
\newcommand{\Smilei}{{\comicneue Smilei}}

\newcommand{\micro}{\text{\textmu}}
\newcommand{\ee}{\mathrm{e}}
\newcommand{\ii}{\mathrm{i}}
\newcommand{\pp}{\mathrm{p}}

\usepackage{etoolbox}
\makeatletter
\patchcmd{\NAT@citex}
{\@citea\NAT@hyper@{%
    \NAT@nmfmt{\NAT@nm}%
    \hyper@natlinkbreak{\NAT@aysep\NAT@spacechar}{\@citeb\@extra@b@citeb}%
    \NAT@date}}
{\@citea\NAT@nmfmt{\NAT@nm}%
  \NAT@aysep\NAT@spacechar\NAT@hyper@{\NAT@date}}{}{}
\patchcmd{\NAT@citex}
{\@citea\NAT@hyper@{%
    \NAT@nmfmt{\NAT@nm}%
    \hyper@natlinkbreak{\NAT@spacechar\NAT@@open\if*#1*\else#1\NAT@spacechar\fi}%
    {\@citeb\@extra@b@citeb}%
    \NAT@date}}
{\@citea\NAT@nmfmt{\NAT@nm}%
  \NAT@spacechar\NAT@@open\if*#1*\else#1\NAT@spacechar\fi\NAT@hyper@{\NAT@date}}
{}{}

\gdef\@underjournal{%
  \vbox to 5.5\p@{\noindent
    \parbox[t]{4.5in}{\normalfont\indexsize Published in: \textit{J. Plasma Phys.}
      (2022), \textit{vol.} 88, 905880211 \quad \textcopyright{} The
      Author(s), 2022.\\
      This is an Open Access article, distributed under the terms of
      the Creative Commons Attribution licence
      (\url{http://creativecommons.org/licenses/by/4.0/}), which
      permits unrestricted re-use, distribution and reproduction,
      provided the original article is properly
      cited. doi:\href{https://doi.org/10.1017/S0022377822000307}{10.1017/S0022377822000307}
      \\[2.5\p@]
      {\ \ }}%
  \vss}%
}
\makeatother

\begin{document}

\shorttitle{Diagnosing plasmas with attosecond dispersion}
\shortauthor{A.~Sundstr\"{o}m, I.~Pusztai,
  P.~Eng-Johnsson and T.~F\"{u}l\"{o}p} %
\title{\vspace{2em}Attosecond dispersion as a
  diagnostics tool for solid-density laser-generated plasmas
} %
\author{Andr\'{e}as Sundstr\"{o}m\aff{1}
  \corresp{\email{andsunds@chalmers.se}}, %
  Istv\'{a}n Pusztai\aff{1}, %
  Per Eng-Johnsson\aff{2} %
  \and T\"{u}nde F\"{u}l\"{o}p\aff{1}
  }
\affiliation{
  \aff{1} Department of Physics, Chalmers University of
  Technology, 412\,96 Gothenburg, Sweden
  \aff{2} Department of Physics, Lund University, 223\,63 Lund, Sweden
}
\maketitle

\begin{abstract}
Extreme-ultraviolet pulses can propagate through ionised solid-density
targets, unlike optical pulses and, thus, have the potential to probe
the interior of such plasmas on sub-femtosecond timescales. We present
a synthetic diagnostic method for solid-density laser-generated
plasmas based on the dispersion of an extreme-ultraviolet attosecond
probe pulse, in a pump--probe scheme.
We demonstrate the theoretical feasibility of this approach through
calculating the dispersion of an extreme-ultraviolet probe pulse
propagating through a laser-generated plasma. The plasma dynamics is
calculated using a particle-in-cell simulation, whereas the dispersion
of the probe is calculated with an external pseudo-spectral wave
solver, allowing for high accuracy when calculating the dispersion.
The application of this method is illustrated on thin-film plastic and
aluminium targets irradiated by a high-intensity pump pulse. By
comparing the dispersion of the probe pulse at different delays
relative to the pump pulse, it is possible to follow the evolution of
the plasma as it disintegrates. The high-frequency end of the
dispersion provides information on the line-integrated electron
density, whereas lower frequencies are more affected by the highest
density encountered along the path of the probe. In addition, the
presence of thin-film interference could be used to study the
evolution of the plasma surface.
\end{abstract}

\section{Introduction}

The interaction between high-intensity lasers and solids has many
promising potential applications, such as ion acceleration
\citep{Romagnani-etal_PRL2005,Zhang-etal_PRL2017,
  Higginson-etal_NComm2018}, %
warm-dense-matter generation %
\citep{Remington-etal_Sci1999,Renaudin-etal_PRL2003,
  Perez-etal_PRL2010,Brown-etal_PRL2011}, %
and inertial fusion \citep{Drake_NF2018,LePape-etal_PRL2018}.
It is therefore of great importance to further our understanding of
these interactions. Insight may be gained through both experiments and
numerical simulations, but in order to maximise the utility of the two
tools, we must also be able to combine them through experimental
diagnostics and comparisons with simulations.

In experiments using solid-density laser-generated plasmas, we have a
limited number of experimental diagnostics methods, such as x-ray
radiation \citep{Chen-etal_pop2009,Renner-Rosmej_MRE2019} or ejected
particles
\citep{Neely-etal_AppPhysLett2006,Nurnberg-etal_RevSciInstr2009}.  In
addition to experimental diagnostics, numerical simulations~--~in
particular using the particle-in-cell (PIC) method~--~are widely used
to gain understanding of the evolution of the laser-generated plasma.
However, for PIC codes the particle noise is an inherent source of
error, and practical limitations on the resolution can lead to
unphysical behaviour in certain cases \citep{Juno-etal_JPP2020}.  In
addition, at solid density, especially when the Coulomb logarithm is
order unity, the classical two-body treatment of collisions breaks
down \citep{Starrett_PoP2018}, thus collision models rely on
\textit{ad hoc} choices and, especially because different choices may
lead to significantly different predictions \citep{IonPaper2020}, they
need to be validated experimentally.
Thus, in order to validate numerical simulations of such high-density
laser-generated plasmas, there is a need for additional experimental
diagnostics methods that can be used for validation in this density
regime.

On the experimental side, there are, for instance, optical probing
methods, e.g.\ shadowgraphy, which have been successful in diagnosing
laser-generated plasmas in high detail
\citep{Savert-etal_PRL2015,Siminos-etal_PPCF2016}. However, optical
probing methods are limited to low-density plasmas, typically gas-jet
targets, due to the plasma transparency limit. %
To probe the interior of plasmas at solid density,
\citet{Kluge-etal_PRX2018} employed the scattering of x-rays from a
free-electron laser to study the density evolution in solid density
plasma gratings. With the advent of high-harmonic generation
\citep{Ferray-etal_JPB1988}, attosecond extreme-ultraviolet (XUV)
pulses are more readily available to small-scale labs. %
Furthermore, with reliable methods of generating and measuring XUV
pulses \citep{Calegari-etal_JPB2016,Koliyadu-etal_Photonics2017},
employing XUV frequencies in optical probing methods is now becoming a
possibility, for instance, %
spectral interference methods on the spectral fringes of an attosecond
pulse train \citep{Salieres-etal_PRL1999,Descamps-etal_OptLett2000,
  Merdji-etal_LasPartBeams2000,Hergott-etal_JOSAB2003}, %
high-harmonic transmission spectroscopy to measure electron density
\citep{Hergott-etal_LasPartBeams2001,Doboz-etal_PRL2005}, %
and measurement of XUV refractive index in solid-density plasmas
\citep{Williams-etal_PoP2013}.

In this paper, we present a method based on the \emph{dispersion} of
an attosecond XUV pulse (or a pulse-train) to diagnose plasmas that
are over-dense at optical frequencies. We employ a linearised
pseudo-spectral (PS) wave solver to compute the dispersion of the
probe pulse for any given plasma profile. In particular, we extract
the spatiotemporal plasma profile information along the optical axis
from a PIC simulation, through which the probe pulse is propagated
separately using the PS solver. %
The effect of the probe pulse on the plasma evolution is thus
neglected. This limitation may be possible to relax if necessary,
through a PS fluid modelling of the plasma
\citep{Siminos-etal_PRE2014}, but because experimentally available XUV
intensities are quite low (normalised relativistic amplitudes
$a_1\lesssim10^{-3}$), our linearised treatment of the probe pulse is
justified.
The computed pulse dispersion~--~our synthetic diagnostic~--~is most
sensitive to the electron density variation across the target, whereas
corrections related to the energy distribution of electrons might
provide additional constraints at high electron temperatures.

Utilising such modelling in comparison with experimental measurements
of the dispersion of the high-frequency pulse across the target,
e.g.~with the RABBIT \citep{Paul-etal_Sci2001} or attosecond
streak-camera \citep{Itatani-etal_PRL2002} methods, could provide an
experimental diagnostic. Indeed, this type of experiment have already
been performed with (non-ionised) aluminium foil by
\citet{LopezMartens-etal_PRL2005}, but not yet in comparison with
simulations of the plasma evolution. In addition, the use of isolated
attosecond probe pulses in our study provides unprecedented temporal
resolution. %
We note, however, that the use of an isolated pulse is not strictly
necessary~--~although it simplifies the numerical analysis. The main
challenge of employing a pulse train, instead of a single pulse, stems
from the spectral fringes that it produces, as discussed in
\S\,\ref{sec:results}.

The structure of this paper is as follows. In \S~\ref{sec:dispersion}
we derive the plasma dispersion for a low-amplitude wave in the
presence of relativistic electrons with arbitrary momentum
distribution. This is followed in \S~\ref{sec:PS} by a description of
a linearised PS wave solver, which allows computation of
the group delay of the frequency components in a probe pulse, with
minimal numerical dispersion. This tool set is then applied to an
output from a PIC simulation in \S~\ref{sec:pic}. The results are
presented and discussed in \S~\ref{sec:results} and the conclusions
summarised in \S~\ref{sec:concl}.

\section{Dispersion of the probe pulse}
\label{sec:dispersion}

Our goal is to model the propagation of a high-frequency pulse in a
laser-generated plasma. Although a PIC code is suitable to model the
dynamics of the plasma and the laser pulse used to generate it, there
are reasons to model the evolution of the high-frequency probe pulse
in an external numerical tool, which is the approach we
adopt. Resolving the spatiotemporal scales of the probe within the PIC
code would already require excessive numerical resources, while
maintaining the accuracy of the phase evolution, such that it is
suitable for experimental comparison, in a finite-difference
time-domain framework affected by numerical dispersion
\citep{Nuter-etal_EPJD2014}, is simply not feasible (note that even in
a non-standard, spectral PIC code with lower numerical dispersion, the
extremely high resolution requirement would still persist; there may
also be potential issues with the level of discretisation noise
compared with the probe pulse amplitude.)

In the following, we derive the plasma response to a spatial harmonic
of the probe pulse, in the presence of electrons with arbitrary
momentum distribution $f(\vb*p)$. The resulting expression is then
used in the numerical framework, external to the PIC code, to evolve
the probe pulse based on the plasma information obtained from the PIC
simulation, described in \S\,\ref{sec:PS}.

We treat the probe pulse as a ray propagating along a line in the
plasma, assuming negligible plasma variations over the transverse
extent of the probe pulse, thereby keeping the formalism
one-dimensional (1D). Consider a transverse electromagnetic
wave\footnotemark{}, described by its vector potential
$\vb*{A}_{\perp}$. %
\footnotetext{Although it is possible to generate longitudinal
    wave components in a plasma with relativistic laser pulses
    \citep{Kaw-Dawson_PhysFluids1970}, we will restrict our analysis
    to low-amplitude waves, for which the transverse--longitudinal
    coupling is very weak.}%
In a 1D geometry, conservation of transverse canonical momentum
dictates that the momentum response of the electrons is
$\vb*{p}_{\perp}=e\vb*{A}_{\perp}$, where $-e$ is the electron charge.
In a cold-fluid plasma, assuming stationary ions, the corresponding
current response is
\begin{equation}\label{eq:j_perp}
\vb*{j}_{\perp}=-en_{\ee}\vb*{v}_{\ee,\perp}
=-\frac{e^2n_{\ee}\vb*{A}_{\perp}}{\gamma_{\ee}m_{\ee}},
\end{equation}
where $n_{\ee}$ is the electron density,
$\vb*{v}_{\perp}=\vb*{p}_{\perp}/m_{\ee}\gamma_{\ee}$ is the fluid
velocity of the electrons, $m_{\ee}$ is the electron mass, and
$\gamma_{\ee}$ is the Lorentz factor of the electron fluid motion.
We can now write down the 1D wave equation for a cold-fluid plasma as
\begin{equation}\label{eq:wave}
\pdv[2]{\vb*A_{\perp}}{t} - c^2\pdv[2]{\vb*A_\perp}{x}
= c^2\mu_{0}\vb*{j}_{\perp}
=-\frac{\omega_{\pp}^2}{\gamma_{\ee}}\vb*{A}_{\perp},
\end{equation}
where the wave is propagating along the $x$ direction, $c$ is the
speed of light in vacuum, and
\begin{equation}\label{eq:wp_sq_0}
\omega_{\pp}^2\equiv\frac{e^2n_{\ee}}{\epsilon_{0}m_{\ee}}
\end{equation}
defines the non-relativistic plasma frequency $\omega_{\pp}$.

For a low-amplitude wave, such as the high-harmonic generated pulses
available today, $a_1=eA_1/(m_{\ee}c)\ll1$ where $A_1$ is the
amplitude of the wave vector potential, we may neglect the
contribution from the wave-induced oscillation to the Lorentz factor
$\gamma_{\ee}$. For the moment, we also assume that the longitudinal
fluid momentum of the electrons is small, $p_{x}\ll m_{\ee}c$, such
that $\gamma_{\ee}\simeq1$. Thus, the prefactor on the-right hand side
of \eqref{eq:wave} is independent of $\vb*{A}_{\perp}$, and we recover
the cold-plasma dispersion relation
$c^2k^2 = \omega^2-\omega_{\pp}^2$.

Note that in this derivation we have neglected effects of particle
collisions, which, if sufficiently strong, could affect both the phase
shifts and damping of the wave. In the cases considered in this paper,
however, we estimate the electron--ion collision frequencies
$\nu_{\ee\ii}$ to be one to two orders of magnitude lower than the XUV
probe frequencies~--~owing to the approximately kiloelectronvolt
electron temperatures reached~--~and can therefore be neglected (note
that collisional corrections to phase delays are quadratic in
$\nu_{\ee\ii}/\omega\ll 1$). In cases where collisions play a larger
role, e.g.\ in colder plasmas, the wave equation \eqref{eq:wave} could
be modified to accommodate collisional effects via a damping term
$\tilde{\nu}\pdv*{\vb*{A}_{\perp}}{t}$ (added to the left-hand side),
where $\tilde{\nu}$ is an effective damping rate due to
collisions. Finding an appropriate expression for $\tilde{\nu}$ is
non-trivial, but once that is done, the PS solver outlined in
\S\,\ref{sec:PS} is easily modified to include this damping term.

\subsection{Relativistic birefringence: a kinetic correction to the
  plasma frequency}
\label{sec:birefringence}

In the above presentation, we used a cold-fluid description of the
plasma response to the electromagnetic wave.  In reality, the
electrons are not cold, and relativistic effects will require a
modification to the current response of an individual electron based
on its initial momentum, as was pointed out by
\citet{Stark-etal_PRL2015} and further developed by
\citet{Arefiev-etal_PoP2020}. In general, the Lorentz factor
contributes to increasing the inertia of the electrons, which lowers
the effective plasma frequency.
This effect is akin to relativistic transparency
\citep{Kaw-Dawson_PhysFluids1970,Siminos-etal_PRE2012}, but here it is
the general effect of relativistic plasmas~--~not just the
relativistic electron motion induced by the laser. In this subsection,
we present the relativistic, kinetic correction to the plasma
frequency for a low-amplitude wave.

For an electron subject to a field with vector potential
$\vb*{A}_\perp=A_y\vu*{y}$, its corresponding change in momentum would
be $\delta{p_y}=eA_y$ due to conservation of transverse canonical
momentum. Importantly, this change in momentum is \emph{independent of
  the initial momentum of the electron}. However, the velocity
response~--~and thus also the current response~--~does depend on the
initial momentum. Therefore, when calculating the velocity response,
we must consider the initial Lorentz factor of the electron,
$\gamma=(1+p_x^2+p_y^2+p_z^2)^{1/2}$. Here, and in the rest of this
subsection, we use the normalisation $m_{\ee}=1=c$, unless stated
otherwise.

Small wave amplitudes allow linearisation of the change in velocity
$\vb*{v}=\vb*{p}/\gamma$ due to the small momentum perturbation
$\delta{p_{y}}$ in the $y$ direction (polarisation direction). The
corresponding change in velocity can be expressed as
\begin{equation}
\delta{v_y}\approx\delta{p_y}\dv{v_{y}}{p_{y}}
=\delta{p_y}\qty(\pdv{v_{y}}{p_{y}} +
\pdv{v_{y}}{\gamma}\pdv{\gamma}{p_{y}})
=\frac{\delta{p_y}}{\gamma}\qty(1-\frac{p_y^2}{\gamma^2}).
\end{equation}
From this expression, we can note that the corresponding change in
velocity depends on the initial momentum and the respective direction of
the momentum perturbation.

Next, we obtain the full current response of the plasma by integrating
the individual velocity response $\delta{v_y}$ over the electron
distribution function $f$,
\begin{equation}
\delta{j}_{y}=-e\int\!\dd[3]{p}\,\delta{v_y}f(\vb*p)
=-e\,\delta{p_{y}}
\int\!\dd[3]{p}\frac{f(\vb*p)}{\gamma}\qty(1-\frac{p_y^2}{\gamma^2}).
\end{equation}
In particular, we may lift the momentum perturbation
$\delta{p}_y=eA_{y}$ outside the integral because the conservation of
canonical momentum applies to each electron, regardless of its initial
momentum.
In the previous calculations, we have used $\vu*{y}$ as the direction of
polarisation $\vb*A_\perp=A_y\vu*{y}$, but the same calculations can
easily be extended to an arbitrary polarisation direction $\vu*{e}$,
by replacing all instances of $p_y$ with $\vb*{p}\vdot\vu*{e}$. 

Finally, we note that we can obtain the kinetically corrected
dispersion relation by replacing the non-relativistic plasma frequency
\eqref{eq:wp_sq_0} with a \emph{relativistic plasma frequency}
\begin{equation}\label{eq:wp_sq_1}
\tilde{\omega}_{\pp,\vu*{e}}^2=\frac{e^2}{\epsilon_{0}m_{\ee}}
\int\!\dd[3]{p}\frac{f(\vb*p)}{\gamma}
\qty(1-\frac{(\vb*{p}\vdot\vu*{e})^2}{\gamma^2})
=\frac{e^2n_{\ee}}{\epsilon_{0}\tilde{\gamma}_{\vu*e}m_{\ee}},
\end{equation}
where $m_{\ee}$ has been added back to the prefactor for clarity.  The
relativistic plasma frequency can then simply be used in the
plasma-response term in the wave equation \eqref{eq:wave}. In the last
step, we also introduced an \emph{effective gamma factor}, defined by
\begin{equation}\label{eq:gamma-tilde}
\tilde{\gamma}^{-1}_{\vu*{e}}\equiv\frac{1}{n_{\ee}}
\int\!\dd[3]{p}\frac{f(\vb*p)}{\gamma}
\qty(1-\frac{(\vb*{p}\vdot\vu*{e})^2}{\gamma^2}).
\end{equation}
The effective gamma factor corresponds to the relative difference
between the relativistic and non-relativistic plasma frequencies,
$\tilde{\gamma}^{-1}_{\vu*{e}}=\tilde{\omega}_{\pp,\vu*{e}}^2/\omega_{\pp}^2$.
If the distribution is not isotropic, then
$\tilde{\gamma}_{\vu*{e}}$~--~and thereby
$\tilde{\omega}_{\pp,\vu*{e}}^2$~--~are polarisation dependent, hence we
can talk about this effect as a ``relativistic
\emph{birefringence}''\footnotemark{}. %
\footnotetext{This nomenclature follows \citet{Arefiev-etal_PoP2020}.
  However, the same term has also been used before by
  \citet{Schwab-etal_PRAB2020}, although for a different physical
  phenomenon: the birefringence reported in their paper is due to
  strong magnetic fields. }

For thermal distributions at temperatures below approximately
$10{\rm\,keV}$, we show in Appendix~\ref{sec:thermal}, that the
relative reduction of the plasma frequency due to the effective gamma
factor is only of the order of a few per cent. Therefore, in most
experimentally relevant cases, the attosecond dispersion can be said
to probe the electron density.

\section{Linearized pseudo-spectral wave solver}
\label{sec:PS}

A spectral solver is based on discretising and evolving the wave
spectrum rather than the real-space wave. The main benefit of a
spectral solver is that numerical dispersion can be reduced
drastically compared with spatial discretisations, such as the
commonly used Yee mesh. Although numerical dispersion in the context
of PIC codes \citep{Nuter-etal_EPJD2014} is often discussed with
respect to numerical Cherenkov radiation
\citep{Godfrey_JCompPhys1974}, it can also greatly affect the accuracy
of the evolution of the electromagnetic wave, which is the main
problem addressed here.

In a periodic domain of length $L$, we may Fourier decompose
the wave vector potential in space into spectral modes,
\begin{equation}
\vb*{A}_{\perp}(x;t) = \sum_{k}\vb*{\hat{A}}_{\perp,k}(t)\,\ee^{\ii kx},
\end{equation}
where the sum is over all integer multiples of $\rmDelta{k}=2\upi/L$;
once the domain has been discretised into $N$ equally spaced points,
the upper and lower limit to the sum are
$\pm N\rmDelta{k}/2=\pm N\upi/L$. %
We may also transform
$\tilde{\omega}_{\pp}^2(x;t)\mapsto\hat{\omega}_{\pp,k}^2(t)$
analogously. Through this Fourier decomposition, we can replace the
spatial derivatives $\pdv*{x}\mapsto\ii{k}$, and the wave equation
\eqref{eq:wave} now becomes
\begin{equation}\label{eq:wave_k-sum}
\pdv[2]{\vb*{\hat{A}}_{\perp,k}}{t}+c^{2}k^2\vb*{\hat{A}}_{\perp,k}
=-\sum_{k'}\hat{\omega}_{\pp,k'}^2\vb*{\hat{A}}_{\perp,(k-k')}.
\end{equation}
It has been reduced to a set of coupled ordinary differential
equations, which can be solved using standard methods, e.g.\
Runge--Kutta.
The right-hand side of \eqref{eq:wave_k-sum} is a convolution in $k$
space (according to the \textsl{Convolution theorem} of Fourier
analysis) which can be evaluated efficiently using a PS approach,
i.e.\ it is transformed back to real space in each time step, where
this convolution is reduced to a multiplication. This is particularly
convenient as $\tilde{\omega}_{\pp}^2(x;t)$ is provided in real space
from the simulations.

In this method, we have implicitly assumed a periodic domain. For the
purposes of this paper~--~to study the dispersion of an XUV pulse
through a foil-target laser-generated plasma~--~the simulation can be
accommodated in such a periodic domain by allowing for sufficiently
large vacuum regions on both sides of the laser-generated plasma. In
such a simulation, there is no need for the XUV pulse to cross the
periodic boundary.  Other types of simulations, where the plasma
extends the full length of the simulation box and where the XUV pulse
propagates though the periodic boundary can also be accommodated, as
long as the XUV pulse does not cross the boundary of the plasma
simulation~--~e.g.~by projecting a moving window plasma-simulation
domain onto the PS periodic domain.

Finally, we note that the numerical methods for solving the
second-order differential equation usually involves decomposing the
time derivative into two first-order derivatives. By doing so, we
obtain the electric field $\vb*{\vb*{E}}=-\pdv*{\vb*{A}}{t}$~--~which
is the quantity that can be measured in experiments~--~essentially `for
free'.

\subsection{Calculating phase shifts}
\label{sec:phase-shift}
The main result that we seek from the PS computation is the dispersion
or relative phase shifts of the frequency components of the XUV pulse,
which can be measured experimentally. Although the phase $\phi_k(t)$
is encoded in the Fourier spectrum as the complex phase of
\begin{equation}
\hat{E}_k(t) = \abs*{\hat{E}_k(t)}\ee^{\ii\phi_k(t)}
\equiv  \abs*{\hat{E}_k(t)} P_k(t),
\end{equation}
it is not trivial to recover $\phi_k(t)$ from $P_k(t)$, because
$P_k(t)$ only contains phase information modulo $2\pi$. Practically,
this makes it nearly impossible to reconstruct the relative phases of
two frequency components separated by more than a few steps in the
discretised $k$ space.

Instead of analysing $P_k$ directly, we may use
$\bar{P}_k(t)\equiv P_{k}(t)\ee^{\ii{}\omega_{k}t}
=\ee^{\ii[\phi_k(t)+ckt]}=\ee^{\ii\bar\phi_k(t)}$, %
which will make the phase-shift analysis clearer. This view removes
the relative phase shifts due to vacuum propagation, and is equivalent
to studying the pulse in a comoving window that moves with the speed
of light. Any phase shifts observed in this frame is therefore due to
the plasma dispersing the pulse.

As the full information of the relative phase shifts is difficult to
extract directly from the complex spectral phase $\bar{P}_k$, some
other method is necessary. Fortunately, because we are interested in
the \emph{relative} phases, we can study the change of $\bar{P}_k$,
\begin{equation}\label{eq:deltaP}
\begin{aligned}
\rmDelta\bar{P}_k ={}& \bar{P}_{k+1}-\bar{P}_{k}
=\ee^{\ii\bar{\phi}_{k+\rmDelta{k}}}-\ee^{\ii\bar{\phi}_{k}}
=\ee^{\ii\bar{\phi}_{k}}\big[\ee^{\ii\rmDelta\bar{\phi}_{k}}-1\big]\\
={}&\bar{P}_{k}\;
\qty[\ii\sin(\rmDelta\bar{\phi}_{k})+\cos(\rmDelta\bar{\phi}_{k})-1],
\end{aligned}
\end{equation}
where the phase variation is
$\rmDelta\bar{\phi}_{k}=\bar{\phi}_{k+\rmDelta{k}}-\bar{\phi}_{k}$. %
From \eqref{eq:deltaP}, we can define a related phase-rate variable
\begin{equation}\label{eq:psi_k}
\bar{\psi}_k\equiv-\frac{\ii\rmDelta\bar{P}_{k}}{\bar{P}_{k}}
=\sin(\rmDelta\bar{\phi}_{k})+\ii\qty[1-\cos(\rmDelta\bar{\phi}_{k})].
\end{equation}
This variable is still $2\pi$ periodic, but now the periodicity is in
$\rmDelta{\bar{\phi}_{k}}$, which is smaller than the $\pm\pi$ window of
retrievable information~--~provided that the spectral resolution $N$
is sufficiently large.
We can, therefore, retrieve $\rmDelta\bar{\phi}_{k}$, with which the
properly unwrapped phase $\phi_{k}$ can be reconstructed, up to
a constant phase shift.

In an experiment, however, it is the relative \emph{group delay} of
the different frequency components that is measured. The group delay
$\tau$ is defined as the rate of phase change, which, in the discrete
case, we approximate as
\begin{equation}\label{eq:tau_k}
\tau \equiv \pdv{\phi}{\omega} \approx
\frac{\rmDelta\phi_k}{\rmDelta\omega}=
\frac{\rmDelta\phi_k}{c\rmDelta{k}}\equiv\tau_k.
\end{equation}
With this, we now have a full tool set for computing the relative
group delay of a low-amplitude probe pulse in any given 1D plasma
profile $\tilde{\omega}_{\pp}^{2}(x;t)$, with minimal numerical
dispersion. This tool set can be applied on the output form a PIC
simulation to create a synthetic diagnostic for a laser-generated
plasma experiment.

\section{PIC and PS simulations}
\label{sec:pic}
The workflow for generating the synthetic dispersion diagnostic
consists of two main components: first the PIC
simulation to simulate the plasma evolution due to the pump laser
pulse, then the PS method is used to calculate the
dispersion of the probe pulse as it propagates through the plasma
profile generated by the PIC simulation. This two-step process works
for sufficiently low-amplitude probe pulses, where the effects of the
probe pulse on the plasma are negligible.

\subsection{PIC simulation parameters}
We used the PIC code \Smilei{} \citep{Smilei-paper}, to generate
on-axis profiles of the relativistic plasma frequency
$\tilde{\omega}_{\pp}^2(x;t)$ from \eqref{eq:wp_sq_1}, which could
then be used in the linearised PS wave solver. %
To this end, we performed two-dimensional (2D), fully collisional
\citep{Perez-etal_PoP2012}, PIC simulations of a thin plastic foil
target irradiated by a circularly polarised pump pulse with wavelength
$\lambda_0=800{\rm\,nm}$, peak intensity
$1.9{\times}10^{19}{\rm\,W\,cm^{-2}}$ (normalised amplitude
$a_0=3.0$). The pulse temporal and spatial profiles were both Gaussian
with intensity full-width-at-half-maximum (FWHM) duration of $30{\rm\,fs}$
and spot size (waist diameter) of $6{\rm\,\micro{m}}$.%

The thin-foil plasma studied here has a trapezoidal density profile
with a $0.25{\rm\,\micro{m}}$ plateau and $25{\rm\,nm}$ linear
density ramps on both sides, fully ionised, solid-density polyethylene
($\rm CH_2$), corresponding to an initial electron density of
$n_{0,\ee}=177.7\,n_{\rm c,0}=3.1{\times}10^{23}{\rm\,cm^{-3}}$, where
$n_{\rm c,0}=\epsilon_{0}m_{\ee}\omega_{0}^{2}/e^2$ is the critical
density associated with the pump laser frequency $\omega_0$. The
plasma was modelled with 120, 30 and 15 macro-particles per cell for
the electrons, protons and carbon ions, respectively.

The simulation box was $10{\rm\,\micro{m}}$ and $20{\rm\,\micro{m}}$
in the $x$ and $y$ directions, respectively, with $4096$ cells in each
direction, giving a spatial resolution of $\rmDelta{x}_{\rm PIC}=2.44{\rm\,nm}$
and $\rmDelta{y}_{\rm PIC}=4.88{\rm\,nm}$. The target front edge lies at
$x=4.5{\rm\,\micro{m}}$ from the left edge of the box. Peak on-target
intensity occurs at simulation time $t=78.6{\rm\,fs}$.
The simulations output a binned quantity corresponding to the
relativistic plasma frequency squared $\tilde{\omega}^{2}_{\pp,\vu*y}$
\eqref{eq:wp_sq_1}. In these simulations,
$\tilde{\omega}^{2}_{\pp,\vu*y}$ is around $1\,\%$ lower than the
non-relativistic plasma frequency squared \eqref{eq:wp_sq_0}.
Particles within $\pm0.5{\rm\,\micro{m}}$ from the optical axis were
used in the binning, to create the plasma profile used by the PS
solver. The longitudinal resolution of the binning was the same as the
cell length.

We have chosen the target and laser parameters to illustrate one
possible application of the XUV dispersion diagnostics: time-resolving
the disintegration of a thin foil when irradiated by the pump
pulse. As the electrons are energised by the pump laser, they escape
from both the front and back end of the target, which decreases the
density on axis. Therefore, by varying a delay between the probe and
the pump pulse, the different densities can be inferred from the
decrease in dispersion.

As a comparison with the plastic target, we also performed a similar
simulation of a $0.1{\rm\,\micro{m}}$ thin aluminium foil, with
$10{\rm\,nm}$ linear density ramps on each side. The aluminium is
taken to be fully ionised at solid density, corresponding to an
initial electron density of
$n_{0,\ee}=449.4\,n_{\rm c,0}=7.8{\times}10^{23}{\rm\,cm^{-3}}$, which
is very close to $2.5$ times that of the plastic target. The plasma
was modelled with 120 and 40 particles per cell for the electrons and
aluminium ions, respectively. The other parameters were kept the same
as for the plastic target.
The comparison with the thicker plastic target is interesting because
the initial integrated density, along the optical axis, is the same
between the two targets.

\subsection{PS simulation parameters}
\label{sec:PS-params}

The plasma profiles generated in the PIC simulations are used in the
PS solver to accurately and efficiently calculate the dispersion of
the probe pulse. Note that the solver takes both spatial and temporal
variation into account. In order to get the desired resolution, the
PIC output is interpolated in both time and space.

Because of the good numerical accuracy of a spectral solver, the
resolution does not have to be much greater than what was used in the
PIC simulation. The whole length $L_x=10{\rm\,\micro{m}}$ of the
PIC-simulation box was discretised with $N=8196$ points (twice that of
the PIC simulation), corresponding to a resolution of
$\rmDelta{x}_{\rm PS}=1.22{\rm\,nm}$. This discretisation allows for a
maximum wavenumber of $k_{\max}=\pi N/L_x=2.57{\rm\,nm^{-1}}$
(corresponding to a minimum resolved wavelength of
$2\rmDelta{x}_{\rm PS}=2.44{\rm\,nm}$) with a wavenumber resolution of
$\rmDelta{k}=2\pi/L_x=6.28\times10^{-4}{\rm\,nm^{-1}}$. The time
resolution was automatically handled in the Runge--Kutta method
implemented in the function \texttt{solve\_ivp} from \texttt{SciPy}
version \texttt{1.6.3} \citep{SciPy}.%
\footnote{The code package developed for the PS solver,
  as well as tools for extracting and interpolating data from
  \Smilei{} simulations, is freely available at
  \url{https://github.com/andsunds/PseudoSpectral}} %
In Appendix~\ref{sec:benchmark}, we present some benchmarks of the
PS code.

The probe pulse electric field is initialised in real space in the
vacuum region near the front of the target. The initial pulse shape
has a four-cycle-duration $\cos^2$ envelope on a centred sinusoidal carrier
wave. The central wavelength is $\lambda_1=30{\rm\,nm}$ corresponding
to a wavenumber of $k_1=0.21{\rm\,nm^{-1}}$. Note, however, that the
shape of the initial pulse is not very important to the dispersion
measurement.
The pump--probe delay is set by a temporal shift of the
$\tilde{\omega}_{\pp}^2(x;t)$ profiles from the PIC simulation. When
analysing a spectrum from the simulation, it is important that the
whole pulse is in vacuum, so that the wavenumber and the frequency
spectra are the same. Otherwise, distortions of the spectrum due to
the plasma dispersion makes comparisons of spectral data difficult.

With the carrier wavelengths above, the initial target electron
densities correspond to $n_{0,\ee}\approx0.25\,n_{\rm c,1}$ and
$\approx0.6\,n_{\rm c,1}$ for the plastic and aluminium targets,
respectively, where $n_{\rm c,1}$ is the critical density for the
central frequency $\omega_1=ck_1$ of the probe pulse. As these
values are close to unity, there will be a significant portion of the
pulse that is reflected. Because the reflected part propagates in the
opposite direction, it has a very strong influence on the phase
information of the spectrum. It is, therefore, important to filter out
the reflected pulse, and only study the spectrum of the transmitted
part of the pulse. This filtering is most readily done in real space.

As the PS solver evolves each wavenumber component independently, the
group delay incurred for each $k$ component is, in principle, not
affected by the initial pulse shape. (There may be some minor effects
due to the timing of each component in relation to the plasma
evolution, but they would be in order of the ratio of plasma-evolution
rate to pulse duration.) However, the initial pulse shape has one
important effect on the dispersion analysis. That is the initial
relative group delays. With the choice of a symmetric envelope and
symmetric carrier phase, each wavenumber component will have the same
vacuum-propagation-corrected spectral phase $\bar{P}_k=0$, i.e.\ there
are no relative group delays in such a pulse. Conversely, if some
other probe-pulse shape is used, the observed group delays will be
affected by the initial spectral phases of the pulse. For the purposes
of a synthetic diagnostic to an experiment, however, one could use the
same pulse shape as used in the experiment; the resulting group
delay information can then be compared directly with observations.

\section{Results and discussion}
\label{sec:results}
\begin{figure}
\centering
\includegraphics{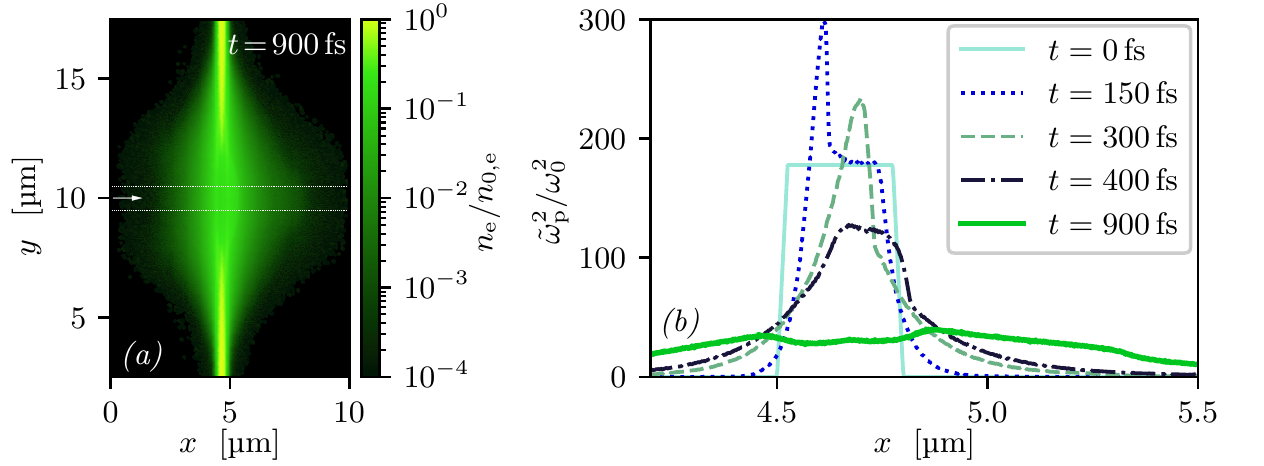}
\caption{(a) Electron density in the polyethylene plasma at
  $t=900{\rm\,fs}$; peak on-target pump intensity occurs at
  $t\approx80{\rm\,fs}$. (b) Profiles of the squared relativistic
  plasma frequency $\tilde\omega_{\pp}^2$, relative to the squared
  pump-laser central frequency $\omega_0^2$, along the optical axis
  (white arrow in panel~a), for $t=0{\rm\,fs}$ (solid line),
  $t=150{\rm\,fs}$ (dotted line), $t=300{\rm\,fs}$ (dashed line),
  $t=400{\rm\,fs}$ (dash-dotted line), and $t=900{\rm\,fs}$ (thick
  solid line). The profiles are created by averaging (in $y$) across a band
  $\pm0.5{\rm\,\micro{m}}$ from the optical axis (thin dotted lines in
  panel~a).  }
\label{fig:PE_dens}
\end{figure}

\begin{figure}
\centering
\includegraphics{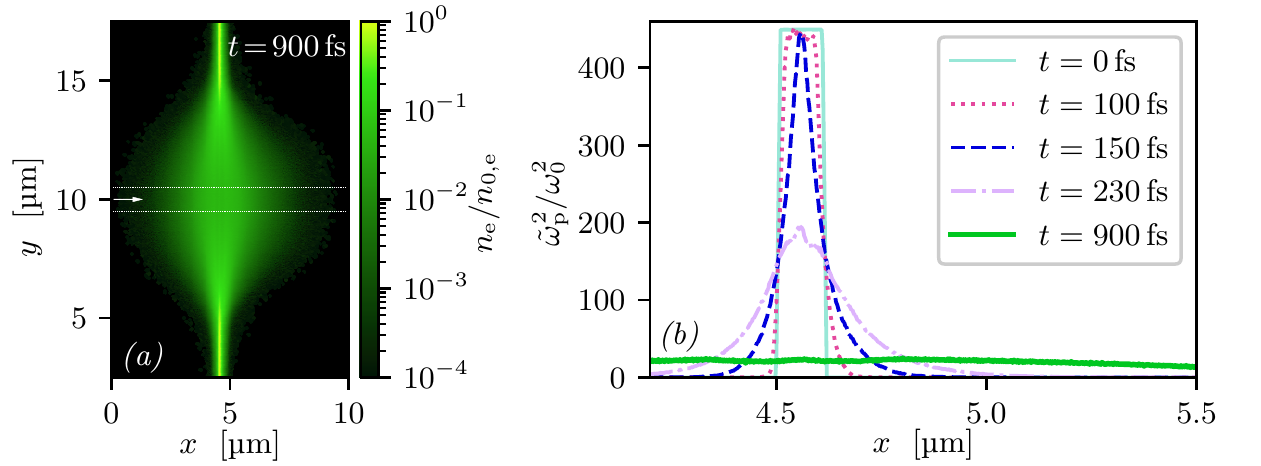}
\caption{(a) Electron density in the aluminium plasma at
  $t=900{\rm\,fs}$; peak on-target pump intensity occurs at
  $t\approx80{\rm\,fs}$. (b) Profiles of the squared relativistic
  plasma frequency $\tilde\omega_{\pp}^2$, relative to the squared
  pump-laser central frequency $\omega_0^2$, along the optical axis
  (white arrow in panel~a), for $t=0{\rm\,fs}$ (solid line),
  $t=100{\rm\,fs}$ (dotted line), $t=150{\rm\,fs}$ (dashed line),
  $t=230{\rm\,fs}$ (dash-dotted line), and $t=900{\rm\,fs}$ (thick
  solid line). The profiles are created by averaging (in $y$) across a
  band $\pm0.5{\rm\,\micro{m}}$ from the optical axis (thin dotted
  lines in panel~a).  }
\label{fig:Al_dens}
\end{figure}

With such thin foils, the targets rapidly disintegrate due to
hydrodynamical expansion after irradiation by the pump
pulse. Figure~\ref{fig:PE_dens}(a) shows an electron-density map, from
the plastic target, at the simulation time $t=900{\rm\,fs}$,
approximately $820{\rm\,fs}$ after peak on-target pump intensity. In
the figure, we see how the electrons have expanded out in plumes, both
in front of and behind the target. Figure~\ref{fig:PE_dens}(b) shows
the on-axis squared relativistic plasma frequency
$\tilde\omega_{\pp}^2$, given by \eqref{eq:wp_sq_1}, for four
different time steps of the simulation. Note that
$\tilde\omega_{\pp}^2$ is approximately proportional to $n_{\ee}$. %
It is these profiles that are being probed by the XUV pulse, which
passes through the plasma along the optical axis
$y=10{\rm\,\micro{m}}$ (white arrow in fig.~\ref{fig:PE_dens}a); the
$\tilde\omega_{\pp}^2$ values are averaged across
$\pm0.5{\rm\,\micro{m}}$ in $y$ on each side of the axis (thin dotted
lines in fig.~\ref{fig:PE_dens}a).

In the evolution of the plasma, it is initially compressed
($t=150{\rm\,fs}$, dotted line) by the pump pulse, creating a density
spike at the front. As the electrons are rapidly heated to
kiloelectronvolt temperatures\footnotemark{}, the plasma starts to
expand and %
expansion fronts (the interface between the perturbed and unperturbed
plasma) propagate inwards from both ends of the target and meet in the
middle. %
\footnotetext{We note that for the parameters considered in this
  paper, collisionless heating mechanisms are sufficient to heat the
  plasma to a temperature comparable to those observed in the
  collisional simulations presented here; this is unlike the
  simulations using heavier target materials and 1D geometry
  \citep{ElectronPaper2020}.}%
As the expansion fronts collide, at first a narrow density peak is
created ($t=300{\rm\,fs}$, dashed line), but that peak is rapidly
flattened as material is continuously escaping on both sides
($t=400{\rm\,fs}$, dash-dotted line). Finally, the plasma continues
to expand, which results in a lower but elongated plasma profile
($t=900{\rm\,fs}$, thick green solid line, corresponding to the
density map in fig.~\ref{fig:PE_dens}a).

A similar evolution is observed in the simulation with the aluminium
foil target. %
Figure~\ref{fig:Al_dens}(a) shows an electron-density map
at $t=900{\rm\,fs}$ (same as the corresponding density map in
fig.~\ref{fig:PE_dens}a). 
The plasma profiles in fig.~\ref{fig:Al_dens}(b) %
(averaged across $\pm0.5{\rm\,\micro{m}}$ on each side of the optical
axis, white dotted lines in fig.~\ref{fig:Al_dens}a) %
show the same general steps in the evolution of the aluminium plasma
as in the plastic-target plasma. Although the initial laser
compression in the aluminium plasma is essentially non-existent
($t=100{\rm\,fs}$, dotted line), there is still the interaction of the
two expansion fronts. At $t=150{\rm\,fs}$, the expansion fronts meet
in the middle, which creates at sharp peak (dashed line). Then, as the
electrons continue to escape, the peak is flattened ($t=230{\rm\,fs}$,
dash-dotted line).

From both fig.~\ref{fig:PE_dens}(a) and \ref{fig:Al_dens}(a), we see
that the transverse variations occur on length scales on the order of
micrometres. Therefore, the transverse beam profile of the probe pulse
should be smaller than that for the 1D assumption in the PS
computation of the probe-pulse dispersion to hold.
\citet{CoudertAlteirac-etal_ApplSci2017} demonstrated an XUV spot size
of ${\simeq}4{\rm\,\micro{m}}$, which is slightly larger than what
would be ideal for the cases studied here. More recently,
\citet{Major-etal_Optica2021} demonstrated a tightly focused waist
(diameter) of $\simeq0.7{\rm\,\micro{m}}$. However, the probe beam
cannot be too tightly focused for the 1D assumption to hold
either. Thus, to avoid the 1D treatment from breaking down, the best
option would be to study plasmas generated by a wider pump pulse, so
that the transverse variations can accommodate a
several-micrometer-wide XUV probe pulse. %

Another possibility to handle transverse plasma variations is to
extend the numerical treatment in the PS solver to two or three
dimensions. Although, the conceptual steps to extend the algorithm to
higher dimensions are simple, some additional complexity arises. The
periodic boundary conditions in the transverse directions~--~caused by
the spectral treatment~--~would have to be handled carefully. In
addition, as the pulse would be dispersed differently in the
transverse plane, the analysis of the arising range of group delays
need to involve a more accurate model of the experimental setup and
detection in the RABBIT or streak-camera methods. We have, therefore,
limited the scope of this study to a 1D dispersion analysis.

\begin{figure}
\centering
\includegraphics{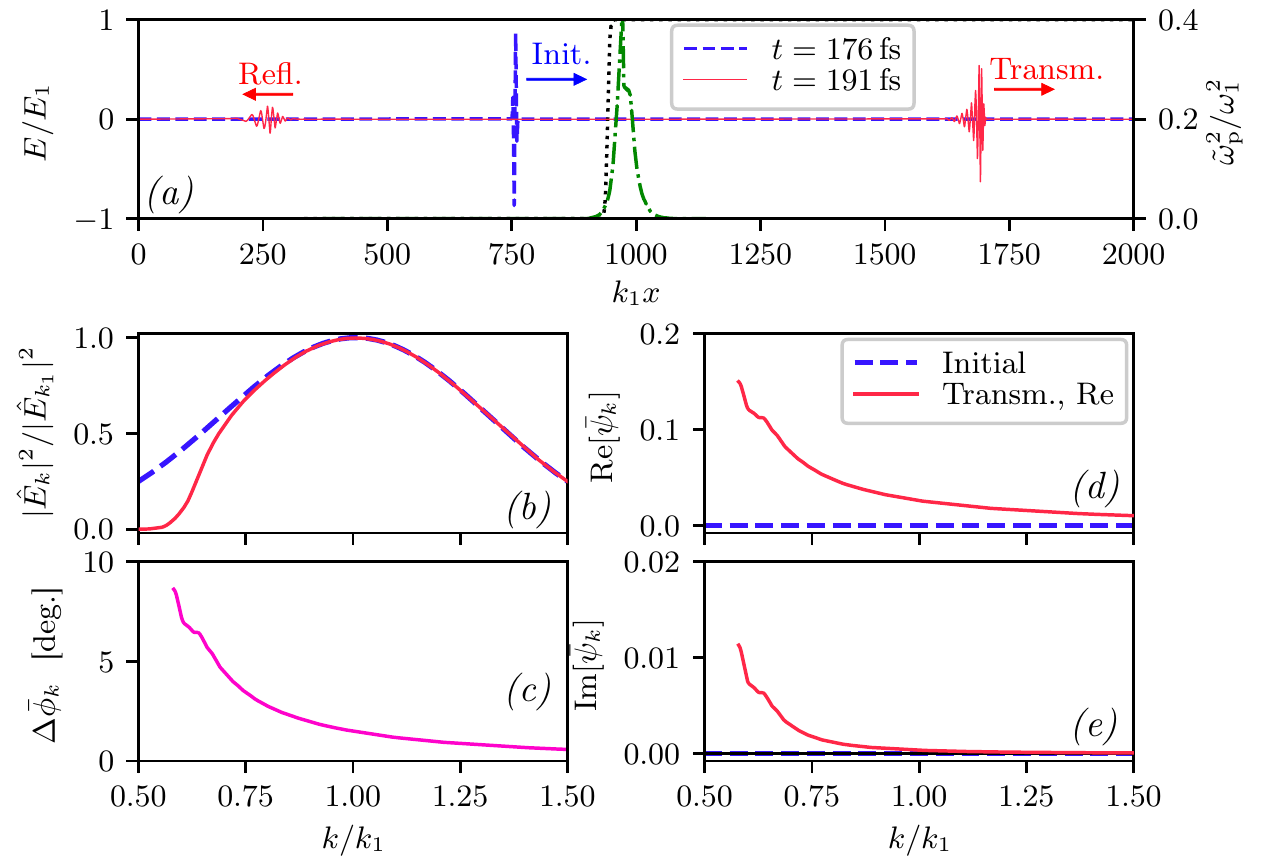}
\caption{Overview of the PS output for the baseline 2D PIC simulation,
  where the XUV probe pulse is $100\rm\,fs$ delayed after the pump. %
  (a) Real-space representation of the probe pulse (blue and red
  lines) and the plasma profile at the final time (green dotted line;
  right axis). The dotted line represents a spatial filter used on the
  reflected part of the pulse when computing the transmitted
  spectra. %
  (b) Initial (dashed line) and transmitted (solid line) spectral
  intensity $|\hat{E}_k|^2/|\hat{E}_{k_1}|^2$ of the probe pulse
  normalised to the maximum value at the central wavenumber $k_1$. %
  (c) Phase variation $\rmDelta\bar{\phi}_k$ of the transmitted
  pulse. %
  (d) Real and (e) imaginary part of the phase-rate variable
  $\bar{\psi}_k$ for the transmitted (solid line) and the initial
  (dashed line) probe pulse. %
}
\label{fig:overview}
\end{figure}

With plasma profiles from a PIC simulation, the attosecond probe pulse
dispersion can now be calculated using the PS wave solver.
Figure~\ref{fig:overview} shows an overview of the information
obtained by such a PS computation, in this case with the plastic
target. In fig.~\ref{fig:overview}(a), the normalised real-space
waveforms of the initial probe pulse at $k_1x=750$ (dashed blue line)
and the final waveform of the transmitted and reflected parts of the
pulse (solid red line), at $k_1x\approx1700$ and $k_1x\approx250$,
respectively. %
We clearly see that the transmitted part of the pulse is significantly
wider than the initial pulse, showing that the pulse has been
dispersed. The reflected pulse is discarded from the later spectral
analysis, via a spatial filter on the real-space waveforms
(represented as the black dashed line in
fig.~\ref{fig:overview}a\footnotemark{}).
\footnotetext{The values of the filter function are not accurately
  represented on any of the axes; the values range from 0 at
  $k_1x\lesssim900$ to 1 at $k_1x\gtrsim1000$, with a sigmoid
  transition in between.}%
The plasma profile ($t=191{\rm\,fs}$) which the pulse
passes through is shown on the right axis relative to the probe
central frequency $\tilde{\omega}_{\pp}^2/\omega_{1}^{2}$ (dash-dotted
green line).

Figure~\ref{fig:overview}(b) shows the normalised energy spectral
density $|\hat{E}_{k}|^2/|\hat{E}_{k_1}|^2$ of the initial (dashed
blue line) and transmitted (solid red line) probe pulse. There is a
clear cutoff in the transmitted spectrum at slightly above the initial
plasma frequency of $\tilde\omega_{\pp,0}=0.50\,\omega_1$.
This is expected, since bulk of the plasma is still at its initial
density\footnotemark{}, making it over-dense for frequencies less than
$\tilde\omega_{\pp,0}=0.50\,\omega_1$. %
\footnotetext{Note that the plasma profile displayed in
  fig.~\ref{fig:overview}(a) is at a later time. The reflection occurs
  at an earlier time, where the plasma profile was closer to its
  initial shape. }

The dispersion of the probe pulse is encoded in the spectral phase
information, which can be retrieved using the methods discussed in
\S\,\ref{sec:phase-shift}. %
Figures~\ref{fig:overview}(d\,\&\,e) show the real and imaginary
parts, respectively, of the phase-rate variable $\bar{\psi}_k$ for the
transmitted pulse as well as the initial pulse. We see that
$\Re(\bar{\psi}_k)=\sin(\rmDelta\bar{\phi}_{k})$ increases as $k$
approaches the cutoff near $k\simeq0.6\,k_1$, which is expected as the
group velocity decreases for frequencies closer to the plasma
frequency. %
Note that the flat initial value of $\bar{\psi}_k=0$ is due to the
choice of initial pulse shape; see \S\,\ref{sec:PS-params} for a short
discussion on the effects of the initial pulse shape. From
$\bar{\psi}_k$, we reconstruct the phase variation
$\rmDelta\bar{\phi}_k$ (fig.~\ref{fig:overview}d) that is proportional
to the group delay, which is what would be measured with the
attosecond streak-camera \citep{Itatani-etal_PRL2002} method in an
experiment.

We have employed an isolated attosecond pulse for the dispersion
analysis in this paper, which somewhat simplifies the computation and
analysis of the group delays. However, it is experimentally more
challenging to generate such isolated attosecond pulses compared to
trains of attosecond pulses. The PS solver and the methods used for
computing the group delays are general and, thus, capable of handling
any waveform~--~including pulse train. However, handling the fringes in
the spectra caused by pulse trains requires some care when analysing
the group-delay data.  Only the group delays at each spectral maxima
should be considered; this corresponds to the information gathered
using the RABBIT \citep{Paul-etal_Sci2001} method. Furthermore, the
plasma processes considered in this paper occur on approximately
$1{-}10{\rm\,fs}$ time scales, which means that the sub-femtosecond
temporal resolution provided by an isolated attosecond pulse is not
strictly necessary.

\subsection{Diagnosing plasma evolution}
\label{sec:plasma-diagnose}

\begin{figure}
\centering
\includegraphics{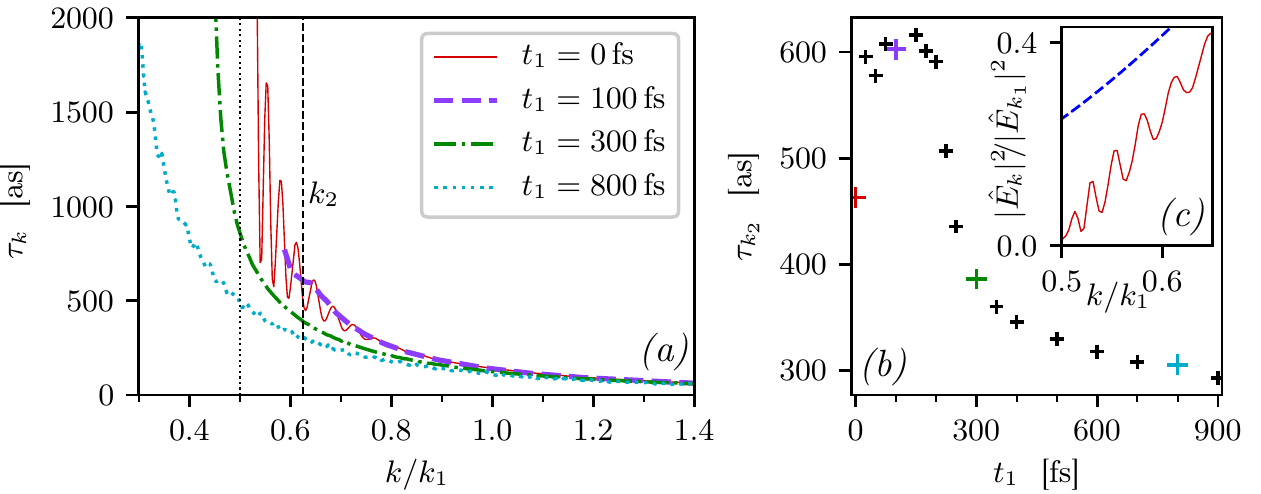}
\caption{(a) Spectral relative group delay $\tau_{k}$ for the plastic
  target as a function of wavenumber $k$ of an XUV probe pulse with
  pump--probe delays $t_1=0$, $100$, $300$ and $800\rm\,fs$ between
  peak intensities of the two pulses. The vertical dotted line
  corresponds to the plasma frequency at the original target density.
  (b) Relative group delay $\tau_{k_2}$ at the wavenumber
  $k_2=0.625\,k_1$ (vertical dashed line in panel a) as a function of
  pump--probe delay $t_1$. Coloured markers correspond to the
  same-coloured curves plotted in panel (a). %
  (c) Spectral intensity $|\hat{E}_k|^2/|\hat{E}_{k_1}|^2$ of the
  initial (dashed blue line) and transmitted (solid red line) probe
  pulse for the $t_1=0{\rm\,fs}$ case.  }
\label{fig:tau}
\end{figure}

Using the group delay as a tool, we can now diagnose the plasma
evolution in the PIC simulations by comparing the group delay for
different pump--probe delays. Figure~\ref{fig:tau}(a) shows the
relative group delay curves $\tau_{k}$ obtained from the plastic
plasma, at four different pump--probe delays $t_1$, where $t_1$ is
measured between the peak intensities of the pump and probe
pulses.
The $\tau_{k}$ curves are plotted in the wavenumber range where the
spectral intensity $|\hat{E}_k|^2/|\hat{E}_{k_1}|^2>0.05$ is greater
than 5\% of that of the central frequency, to remove the
experimentally not measurable regions.
We see that the slope of the $\tau_{k}$ curves decreases as the
pump--probe delay is increased, which means that the dispersion of the
probe pulse decreases. The lowered dispersion is a sign that the
bulk $\tilde{\omega}_{\pp}^2$ decreases with time, as one would
expect from a disintegrating plasma and indeed is observed in
fig.~\ref{fig:PE_dens}(b). We also note that even if the plasma were
to expand one-dimensionally, i.e.\ having a constant line-integrated
density, the $\tau_k$ curves would still be different as changes in
the maximum density would translate to shifts in the cutoff frequency.

Another interesting feature seen in fig.~\ref{fig:tau}(a) is the
oscillation in $\tau_k$ for $t_1=0{\rm\,fs}$ (and a few very weak
oscillations for $t_1=100{\rm\,fs}$). These oscillations occur due to
internal reflections inside the intact target. The XUV pulse is
reflected against the sharp plasma--vacuum boundaries several times,
thus generating a train of very low-intensity pulses following the
main transmitted pulse, which generates interference~--~much like
thin-film interference. The interference pattern can also be seen as
spectral fringes in the intensity spectrum, shown in
fig.~\ref{fig:tau}(c). The reason why the interference patterns do not
appear at longer pump--probe delays $t_1\gtrsim100{\rm\,fs}$ is due to
the destruction of the sharp plasma--vacuum boundaries as the plasma
expands. %
This is interesting, because the presence of thin-film interference in
an XUV probe could be used to study the evolution of the plasma
surface, given that a sufficiently high-contrast pump pulse is used.
Furthermore, the spacing of the fringes can give some information
about the plasma thickness. Note, however, that the fringe separation
depends both on the plasma thickness and the group velocity in the
plasma, the latter being frequency dependent~--~resulting in a
non-constant fringe spacing in fig.~\ref{fig:tau}(c).

The time evolution of the plasma can also be studied by examining the
group delay at a specific frequency. In fig.~\ref{fig:tau}(b),
$\tau_{k_2}$ at $k_2=0.625\,k_1$ (vertical dashed line in
fig.~\ref{fig:tau}a) is shown for a range of different pump--probe
delays from $t_1=0{\rm\,fs}$ to $900{\rm\,fs}$. Interestingly, the
group delay initially increases and reaches a maximum at
$t_1\approx70{\rm\,fs}$%
\footnote{For reference, $t_1=100{\rm\,fs}$ results in the probe pulse
  reaching the target at a simulation time of
  $t\approx180{\rm\,fs}$.}. %
This increase can be attributed to that the maximum electron density
initially increases due to a compression by the laser pressure, as
seen in fig.~\ref{fig:PE_dens}(b), and so does the corresponding
cutoff frequency $\omega_{\pp,\max}$.
After that, $\tau_{k_2}$ starts to drop, which is consistent with
the decreasing maximum density as the plasma starts to
expand hydrodynamically.

\begin{figure}
\centering
\includegraphics{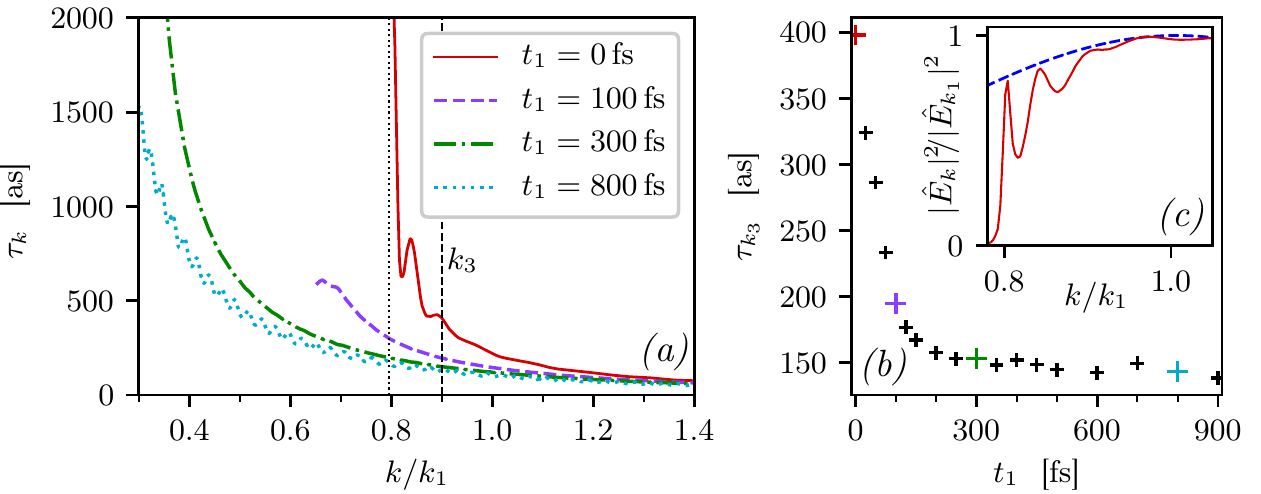}
\caption{(a) Spectral relative group delay $\tau_{k}$ for the
  aluminium target as a function of wavenumber $k$ of an XUV probe
  pulse with pump--probe delays $t_1=0$, $100$, $300$ and $800\rm\,fs$
  between peak intensities of the two pulses. The vertical dotted line
  corresponds to the plasma frequency at the original target density.
  (b) Relative group delay $\tau_{k_3}$ at the wavenumber
  $k_3=0.90\,k_1$ (vertical dashed line in panel a) as a function of
  pump--probe delay $t_1$. Coloured markers correspond to the
  same-coloured curves plotted in panel (a).
  (c) Spectral intensity $|\hat{E}_k|^2/|\hat{E}_{k_1}|^2$ of the
  initial (dashed blue line) and transmitted (solid red line) probe
  pulse for the $t_1=0{\rm\,fs}$ case.
}  
\label{fig:tau_Al}
\end{figure}

Figure~\ref{fig:tau_Al}(a) shows the corresponding relative group
delays in the aluminium plasma. As with the plastic target the
$\tau_{k}$ curves are only plotted in the wavenumber range where
$|\hat{E}_k|^2/|\hat{E}_{k_1}|^2>0.05$. With the aluminium target,
this cutoff moves down in $k$ much faster and farther than in
fig.~\ref{fig:tau}(a), which indicates that the maximum density of the
thinner aluminium drops faster than in the plastic target. In addition
. as in fig.~\ref{fig:tau}(a), the $\tau_k$ curve oscillates for
$t_1=0{\rm\,fs}$ due to the same type of thin-film interference. The
energy spectrum, shown in fig.~\ref{fig:tau_Al}(c), has stronger,
albeit fewer, interference fringes than with the plastic target.

Unlike with the plastic target, there is no initial compression of the
electrons, which results in a monotonically decreasing $\tau_{k_3}$ in
fig.~\ref{fig:tau_Al}(b), for $k_3=0.9k_1$%
\footnote{%
  Owing to the higher initial density of the aluminium target, the
  group delay displayed in fig.~\ref{fig:tau_Al}(b) had to be chosen
  at a higher wavenumber $k_3>k_2$ than in the plastic
  target. Therefore, a direct comparison between $\tau_{k_3}$ in
  fig.~\ref{fig:tau_Al}(b) and $\tau_{k_2}$ in fig.~\ref{fig:tau}(b)
  cannot be done. However, the difference in the qualitative
  behaviours between the two targets is still interesting to study.}.
This observation is consistent with the direct findings from the PIC
simulations shown in fig.~\ref{fig:PE_dens}(b) and
\ref{fig:Al_dens}(b): that the aluminium target lacks a defined
compression wave, and that the inside expansion fronts propagating
from both sides of the plasma meet more rapidly in the thinner target.

Another feature seen in fig.~\ref{fig:tau_Al}(b) is that $\tau_{k_3}$
has an initially steep decay which then changes slope quite abruptly
in a knee at $t_1\simeq150{\rm\,fs}$ (corresponding to a simulation
time of $t\simeq230{\rm\,fs}$, cf.\ fig.~\ref{fig:Al_dens}b). This is
an interesting observation because the knee coincides with the meeting
of the expansion fronts and broadening of the density peaks, which can
be observed directly from the PIC simulation: see the
$\tilde{\omega}_{\pp}^2$ profiles at $t=150{\rm\,fs}$ and
$230{\rm\,fs}$ in fig.~\ref{fig:Al_dens}(b).
A similar, but less-pronounced, knee can also be seen with the
dispersion in the plastic target for $t_1\simeq250{\rm\,fs}$
(simulation time $t\simeq330{\rm\,fs}$). The reason why the knee is
less pronounced in the plastic target is probably that the thicker
plastic target has a wider base density, so that the dispersion is
less dominated by the narrow density peak.

Furthermore, the initial \emph{slope of the decay} of $\tau_{k_3}$ is
correlated with the \emph{rate of decrease} of the peak density. This
relationship can be understood via the group velocity of the plasma
dispersion %
$v_{\rm gr}=c\,(1-\tilde{\omega}_{\pp}^2/\omega^2)^{1/2}$, %
which goes to zero if $\tilde{\omega}_{\pp}$ reaches $\omega$; if we
have a density peak close to the critical density, then the group
delays $\tau_k\sim1/v_{\rm gr}$ near the peak plasma frequency will be
dominated by that peak. %
In the aluminium case shown in fig.~\ref{fig:tau_Al}(b), with $k_3$
quite close to the initial plasma frequency, $\tau_{k_3}$ is first
dominated by the peak density, which is in the form of a narrow peak
that decays rapidly. Later, when the plasma expansion also starts to
flatten the density peak ($t=230{\rm\,fs}$ in
fig.~\ref{fig:Al_dens}b), other contributions to the
dispersion~--~such as the width of the plasma, which evolves more
slowly~--~also become important.
Note, however, that it is more difficult to infer the absolute value
of the peak density, since the absolute value of the group delay will
also be affected by the rest of the plasma~--~by a more slowly varying
additive shift~--~as well as phase shifts when entering/exiting the
plasma.
However, it appears feasible to use XUV dispersion as a direct
diagnostic to infer the evolution of the plasma experimentally.

In the high-wavenumber end of the spectrum, we find another remarkable
feature of the $\tau_k$ curves in figs.~\ref{fig:tau}(a) and
\ref{fig:tau_Al}(a): they all converge in the high-wavenumber
limit. Furthermore, they converge toward the same values for both the
plastic and aluminium targets, e.g.\ $\tau_k\approx40{\rm\,as}$ at
$k=1.8\,k_1$ (outside the ranges displayed in figs.~\ref{fig:tau}a and
\ref{fig:tau_Al}a). This observation can again be understood by
analysing the group velocity in the plasma. The group delay, after
propagation a distance $\ell$, at a frequency $\omega'$ can be
approximated via a Wentzel--Kramers--Brillouin (WKB) approximation (for
slowly varying spatial features and ignoring time variation) as
\begin{equation}
\tau(\omega') +\frac{\ell}{c}
\simeq\int_{0}^{\ell}\frac{\dd{x}}{v_{\rm gr}(x;\omega')}
=\int_{0}^{\ell}\frac{\dd{x}}{c\sqrt{1-\tilde{\omega}^2_{\pp}(x)/{\omega'}^2}}.
\end{equation}
As the group delay, as defined in \eqref{eq:tau_k}, does not take
vacuum propagation into account, a term $\ell/c$ has to be added to the
left-hand side in order to agree with the integral. Note that with the
vacuum propagation accounted for separately in this way, the
integration limits can be chosen arbitrarily large, as long as they
are outside the plasma. Next, in the high-frequency limit, the square
root can be expanded and we obtain
\begin{equation}\label{eq:high-freq-tau}
\tau(\omega'\gg\tilde{\omega}^2_{\pp})\approx
\frac{1}{2c}\int_{0}^{\ell}\frac{\tilde{\omega}^2_{\pp}(x)}{{\omega'}^2}\dd{x}
\simeq\frac{1}{2c}\,\qty(\frac{\omega_1}{\omega'})^2\!
\int_{0}^{\ell}\frac{n_{\ee}(x)}{n_{\rm c,1}}\dd{x};
\end{equation}
we remind the reader that $\omega_1$ and $n_{\rm c,1}$ are the central
frequency and associated critical density of the probe pulse,
respectively. In the last step we also used the approximation
$\tilde{\gamma}\simeq1$, which holds for temperatures
$T_{\ee}\lesssim10{\rm\,keV}$. %
The group delay at high frequencies is therefore a measure of the
line-integrated density, which can be used to follow the transverse
electron transport away from the optical axis during the plasma
expansion.

The fact that all the curves in figs.~\ref{fig:tau}(a) and
\ref{fig:tau_Al}(a) converge toward the same value in the
high-wavenumber limit means that the line-integrated density of the
two plasma on the optical axis, has not changed significantly during
the course of the PIC simulation: recall that the initial
line-integrated densities of the two targets were chosen to be
approximately equal. We may also compare the observed
$\tau_{k'}\approx40{\rm\,as}$ at $k'=1.8\,k_1$
($\omega'=1.8\,\omega_1$) against the known line integrated densities
from the PIC simulations. For the plastic target, the initial density
profile is trapezoidal and can easily be integrated in
\eqref{eq:high-freq-tau} to give a WKB-approximated group delay of
$39{\rm\,as}$. The remarkable agreement between the WKB and PS methods
is a useful sanity check for the PS computation.

In the dispersion analysis of this paper, we have assumed the
amplitude of the XUV probe pulse to be small, which allows for the
linearised treatment of the wave evolution. Although theoretically
convenient, in an experiment, the amplitude of the probe pulse must be
significantly higher than the radiation generated~--~in the relevant
spectral band~--~by the plasma itself, e.g.\ through bremsstrahlung
(BS) and recombination.
Regarding the emission due to recombination, it will be in a limited
number of well-defined wavelengths, which will affect the dispersion
measurements for those specific wavelengths. It should, however, still
be possible to obtain a clean measurements of group delays for the other
XUV wavelengths in the probe-pulse spectrum.

The BS, however, presents a broad-spectrum background noise to the
measurement, thus the XUV probe should have a higher energy than the
detected BS in order to reach a high signal-to-noise ratio.  We
estimate the emitted BS power, in the spectral range down to
$5{\rm\,nm}$ wavelength, to be in the order of
$10^{6}{-}10^{7}{\rm\,W}$ for the plasmas considered in this paper.
However, the BS is emitted isotropically, whereas the probe pulse is
highly directional. The focused XUV pulse reported by
\citet{CoudertAlteirac-etal_ApplSci2017} had an angular divergence of
less than approximately $1{\rm\,mrad}$, corresponding to a solid angle of
$3\times10^{-6}{\rm\,sr}$. Therefore, if the XUV detection is made
with a comparable angular discrimination, the detected power from the
BS would only be in the order of $3{-}30{\rm\,W}$. %
Recently, \citet{Morris-etal_PoP2021} have shown significantly longer
durations of BS emissions than previous literature, up to
$10{-}100{\rm\,ps}$, however that is for much larger target volumes
than considered here. Even with this very conservative estimate of the
duration, the detected BS energy would only be in the
$0.3{-}3{\rm\,nJ}$ range.
In comparison, \citet{Manschwetus-etal_PRA2016} reports an on-target
XUV pulse energy of $40{\rm\,nJ}$, which is sufficiently greater than
the (discriminated) BS power, making group-delay
measurements with the RABBIT or attosecond streak camera methods
feasible. In addition, the signal in these measurements is encoded
as a temporal oscillation with the delay between the XUV pulses and
the external infrared measurement pulse (employed in the RABBIT or streaking
measurement mechanisms), which would further aid to discriminate the
sought signal information from the BS background noise.

\section{Conclusions}
\label{sec:concl}
In this paper, we have presented a synthetic XUV dispersion diagnostic
method, which can be used with the output of a PIC simulation. The use
of XUV frequencies allow for probing of some solid-density plasmas,
which are otherwise over-critical for optical wavelengths. The
synthetic dispersion generated could then be used in comparisons with
experiments, which would aid experimental validation efforts and the
understanding of the evolution of the laser-generated plasma. The
propagation of the XUV probe pulse is accurately calculated using a 1D
PS solver along the optical axis. Then the group delays of the
frequency components of the pulse are computed from the complex phases
of the spectral components.
As a part of the dispersion calculation, we present a linearised
kinetic correction to the plasma frequency, relevant for plasmas that
have a significant fraction of relativistic electrons. However, for
experimentally relevant pump-pulse parameters, this correction only
entails a decrease in effective plasma frequency in the order of a few
per cent.
The main quantity being probed is, therefore, the electron density
profile. To some extent, both the maximum and the line-integrated
value of the density can be inferred from the group delays in the XUV
pulse caused by the dispersion.

We have illustrated this synthetic diagnostic technique on thin foil
targets, where we show that the change in group delays of the XUV
pulse varies significantly for different probe delay times. %
Indeed, the group delays reported here are well within currently
available experimental resolution \citep{LopezMartens-etal_PRL2005}. %
Furthermore, this technique allows for an unprecedented time
resolution of the plasma evolution, which is of great use in
experimental validation of PIC simulations.
This technique might also be used as a direct diagnostic for the
evolution of the peak density in the plasma profile, as well as the
line-integrated density. Furthermore, the presence of thin-film
interference could be used to study the early evolution of the plasma
surface.

\vspace{1em}\noindent%
\textbf{Acknowledgements}\par\vspace{1ex} %
The authors are grateful for fruitful discussions with E.\ Siminos,
L.\ Gremillet, P.\ Tassin, C.\ Riconda, M.\ Hoppe, as well as F.\
P\'{e}rez for support with \Smilei{}.
J.\ Hellsvik at PDC Center for High Performance Computing is
acknowledged for assistance in making \Smilei{} run on the PDC
resources. %

\vspace{1em}\noindent%
\textbf{Funding}\par\vspace{1ex} %
This project has received funding from 
the Knut and Alice Wallenberg Foundation (Dnr.\ KAW\,2020.0111).
The computations were enabled by resources provided by the Swedish
National Infrastructure for Computing (SNIC), partially funded by the
Swedish Research Council through grant agreement no.\ 2018-05973.

\vspace{1em}\noindent
\textbf{Declaration of interests}\par\vspace{1ex} %
The authors report no conflict of interest.

\appendix
\section{Relativistic plasma frequency in thermal plasmas}
\label{sec:thermal}
For an isotropic distribution, i.e.\ $4\pi{}p^2f(\vb*p)=f(p)$, we can
write \eqref{eq:gamma-tilde} as
\begin{equation}
\tilde{\gamma}^{-1}_{\vu*{e}}=
\int\!\dd[3]{p}\frac{\bar{f}(\vb*p)}{\gamma}
\qty(1-\frac{(\vb*{p}\vdot\vu*{e})^2}{\gamma^2})
=\int\!\dd{p}\dd{\theta}\dd{\varphi}\,\sin\theta\,
\frac{\bar{f}(p)}{4\pi\gamma}
\qty(1-\frac{p^2\cos^2\theta}{\gamma^2}),
\end{equation}
where $\bar{f}=f/n_{\ee}$ denote the density-normalised distribution
such that $\int\dd[3]{p}\,\bar{f}(\vb*{p})=1$, and where we have
aligned the $\theta=0$ axis along $\vu*{e}$ direction.
Performing the angular integrals yields the effective gamma factor for
an isotropic distribution as
\begin{equation}\label{eq:gamma-tilde_3d-iso}
\tilde{\gamma}^{-1}
=\frac{1}{3}
\int_{0}^{\infty}\!\dd{p} \frac{\bar{f}(p)}{\gamma}
\qty(2+\frac{1}{\gamma^2})
=\frac{1}{3}
\int_{1}^{\infty}\!\dd{\gamma}
\frac{\bar{f}(\gamma)}{\gamma}
\,\qty(2+\frac{1}{\gamma^2}),
\end{equation}
where in the last integral, we have changed variables to
$\gamma=(1+p^2)^{1/2}$, and we have absorbed the Jacobian into the
distribution function, i.e.\
$\bar{f}(p)\dd{p}=\bar{f}(\gamma)\dd{\gamma}$. This expression for
$\tilde{\gamma}^{-1}$ naturally does not have a polarisation
dependence.

\begin{figure}
\centering
\includegraphics{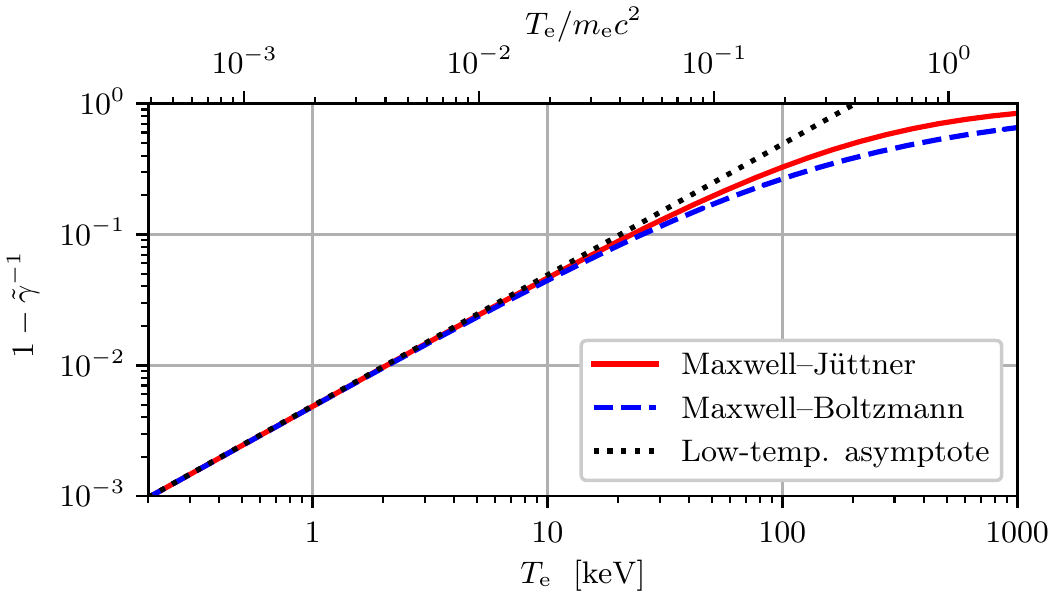}
\caption{Relative reduction of the squared plasma frequency due to
  the effective gamma factor, $1-\tilde{\gamma}^{-1}$, for both
  Maxwell--J\"uttner (solid red line) and Maxwell--Boltzmann (blue
  dashed line) distributed electrons with temperature $T_{\ee}$. The
  curves for both distributions display an asymptotic behaviour
  $1-\tilde{\gamma}^{-1}\simeq5T_{\ee}/(2m_{\ee}c^2)$ as $T_{\ee}\to0$
  (black dotted line). }
\label{fig:gamma-tilde_thermal}
\end{figure}

With \eqref{eq:gamma-tilde_3d-iso}, we can calculate
$\tilde{\gamma}^{-1}$ from a thermal plasma, using the Maxwell--J\"uttner
distribution
\begin{equation}
\bar{f}_{\rm MJ}(\gamma) = \frac{\varTheta}{K_{2}(\varTheta)}\,
\gamma\sqrt{\gamma^2-1}\ee^{-\varTheta\gamma}.
\end{equation}
where $\varTheta=m_{\ee}c^2/T_{\ee}$ is the dimensionless inverse
electron temperature, and $K_2$ is the modified Bessel function of the
second kind (of order two).
If we take this distribution in \eqref{eq:gamma-tilde_3d-iso},
we obtain
\begin{subequations}\label{eq:gamma-tilde_MJ}
\begin{align}
\tilde{\gamma}^{-1}_{\rm MJ}={}&\frac{\varTheta}{3K_{2}(\varTheta)}
\int_{1}^{\infty}\dd{\gamma}\sqrt{\gamma^2-1}\ee^{-\varTheta\gamma}
\,\qty(2+\frac{1}{\gamma^2})\\
={}&\frac{2K_{1}(\varTheta)}{3K_{2}(\varTheta)}
+\frac{\varTheta}{3K_{2}(\varTheta)}
\int_{1}^{\infty}\dd{\gamma}\frac{\sqrt{\gamma^2-1}}{\gamma^2}\ee^{-\varTheta\gamma}.
\end{align}
\end{subequations}
The last integral can be solved numerically, the result of which is
shown in fig.~\ref{fig:gamma-tilde_thermal}.

In order to obtain an explicit analytic result, we can perform the
same calculation for non-relativistic temperatures,
$\varTheta\gg1$. Using the Maxwell--Boltzmann distribution,
\begin{equation}
\bar{f}_{\rm MB}(p)=\frac{4\pi}{(2\pi/\varTheta)^{3/2}}\ee^{-\varTheta\,p^2/2},
\end{equation}
in \eqref{eq:gamma-tilde_3d-iso} yields
\begin{equation}\label{eq:gamma-tilde_MB}
\tilde{\gamma}^{-1}_{\rm MB}=\frac{\varTheta^{3/2}\ee^{\varTheta/4}}{6\sqrt{2\pi}}
\qty[\varTheta K_{0}(\varTheta/4)+(2-\varTheta)K_{1}(\varTheta/4)].
\end{equation}
By asymptotically expanding the Bessel functions in
\eqref{eq:gamma-tilde_MB} for $\varTheta\to\infty$, we find the
low-temperature asymptotic behaviour
\begin{equation}\label{eq:gamma-tilde_M_asymptote}
1-\tilde{\gamma}^{-1}\simeq
\frac{5}{2\varTheta}
=\frac{5T_{\ee}}{2m_{\ee}c^2}
\qas \varTheta\to\infty\quad (T_{\ee}\to0),
\end{equation}
where we have expressed the asymptotic behaviour in terms of the
relative reduction of the squared plasma frequency,
$1-\tilde{\gamma}=1-\tilde{\omega}_{\pp}^2/\omega_{\pp}^2$. Naturally,
\eqref{eq:gamma-tilde_M_asymptote} is also an asymptote to
\eqref{eq:gamma-tilde_MJ} as the Maxwell--J\"uttner distribution
approaches the Maxwell--Boltzmann distribution for low temperatures.

As can be seen in fig.~\ref{fig:gamma-tilde_thermal}, when we
numerically compare the relativistic \eqref{eq:gamma-tilde_MJ} and the
non-relativistic \eqref{eq:gamma-tilde_MB} effective gamma factors, we
find that the two expressions agree well with each other (to within
5\%) up to temperatures of $T_{\ee}\sim10{\rm\,keV}$. In either case,
the relative reduction of the squared plasma frequency,
$1-\tilde{\gamma}=1-\tilde{\omega}_{\pp}^2/\omega_{\pp}^2$, remains
below 5\% for temperatures less than $10{\rm\,keV}$. We also see that the
low-temperature asymptote is a good approximation of the full
solutions, up to temperatures of several kiloelectronvolts.

\section{Benchmarking of the PS solver}
\label{sec:benchmark}

\begin{figure}
\centering
\includegraphics{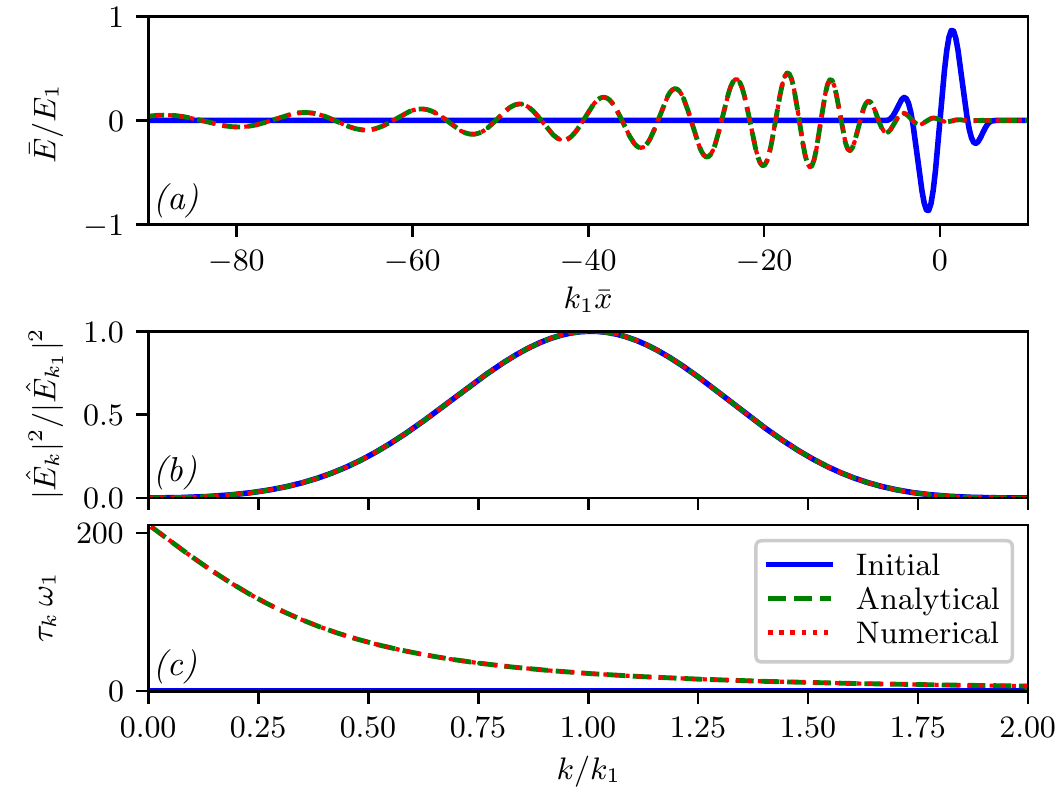}
\caption{Benchmark test of the PS solver: propagation for
  $1{\rm\,\micro{m}}$ through a homogeneous plasma with
  $\omega_{\pp}=0.5\,\omega_1$. A comparison is shown of the
  initial (solid line), analytically (dashed line) and numerically
  propagated (dotted line) probe pulses. %
  (a) Real-space waveforms of the corresponding pulses. The positional
  coordinate $\bar{x}$ is relative to the vacuum propagation position
  of the pulse centre. %
  (b) Energy spectra of the corresponding pulses. %
  (c) relative group delay $\tau_k$ of the numerically and
  analytically propagated pulses. }
\label{fig:benchmark_plasma}
\end{figure}

In order to evaluate the numerical accuracy of the PS solver, we have
benchmarked it by propagating a test pulse through a homogeneous
plasma. The analytical dispersion of a pulse in a homogeneous plasma
with plasma frequency $\omega_{\pp}$ is given by
$\hat{E}_k(t) = \hat{E}_k(t{=}0)\,\exp(-\ii\omega_{k}t),$ where
$\omega_{k} = {\rm sgn}(k)\,(c^2k^2+\omega_{\pp}^2)^{1/2}$ according
to the plasma dispersion. %
Note that, unlike in the main body of the paper, the pulse is
initialised with the waveform shown in
fig.~\ref{fig:benchmark_plasma}(a) \emph{already inside} the plasma,
which means that the wavenumber spectrum, shown in
fig.~\ref{fig:benchmark_plasma}(b), still goes all the way down to
$k=0$ (corresponding to $\omega=\omega_{\pp}$); this is unlike in the
main body of the paper, where the pulse is initialised~--~and then
later measured~--~in vacuum and, therefore, the transmitted part of the
pulse gets cut off below $k_{\text{cutoff}}=\omega_{\pp}/c$.
The benchmark tests were performed with the same numerical settings as
used in the main body of the paper, described in \S\,\ref{sec:PS-params}.

Figure~\ref{fig:benchmark_plasma} shows the results from one such
benchmark, where the pulse has been propagated for
$1{\rm\,\micro{m}}$ through a plasma with plasma frequency
$\omega_{\pp}=0.5\,\omega_1$ (i.e.\ $n_{\ee}=0.25n_{\rm c}$), where
$\omega_1$ is the central frequency of the test pulse. This plasma
frequency is close to that of the plastic target used in this
paper. Figure~\ref{fig:benchmark_plasma}(a) shows the initial (solid
blue line) pulse real-space waveform as well as the analytically
(dashed green line) and numerically (dotted red line) propagated
pulses. There is no discernible difference between the analytical and
numerical cases. Figure~\ref{fig:benchmark_plasma}(b) shows the energy
spectrum, for the three cases as above, and
fig.~\ref{fig:benchmark_plasma}(c) shows the relative group delay
$\tau_{k}=\rmDelta\bar{\phi}_k/\rmDelta\omega$ from
\S\,\ref{sec:phase-shift}. Again, there is no discernible difference
between the numerical and analytical curves; the relative error
between the PS numerical and analytical dispersion, as
represented by $\tau_k$, is less than $10^{-5}$ in the wavenumber
range $0\le{}k\le2\,k_1$.
Similar performances are seen when the plasma density is varied
between $\omega_{\pp}=0$ (vacuum) and $\omega_{\pp}=0.8\,\omega_1$
(aluminium), and with varying propagation lengths up to
$10{\rm\,\micro{m}}$. These benchmark tests demonstrate the extremely
low numerical dispersion of the PS solver.

\bibliographystyle{jpp-like-doi}
\bibliography{references}

\end{document}